\documentclass[twocolumn,aps,superscriptaddress,amsmath,amssymb,nofootinbib,pre]{revtex4}
\usepackage{amssymb,amsfonts,textcomp}
\usepackage{array}
\usepackage{supertabular}
\usepackage{hhline}
\usepackage{graphicx}
\usepackage{xcolor}
\usepackage{soul}

\newcommand{\av}[1]{\left\langle {#1} \right\rangle}

\begin{document}

\title{Synchronization dynamics on the EU and US power grids}

\author{G\'eza \'Odor}
\email[]{odor.geza@ek-cer.hu}
\affiliation{Centre for Energy Research, Institute of Technical Physics\\
and Materials Science, P.~O.~Box 49, H-1525 Budapest, Hungary}
\author{Shengfeng Deng}
\email[]{shengfeng.deng@ek-cer.hu}
\affiliation{Centre for Energy Research, Institute of Technical Physics\\
and Materials Science, P.~O.~Box 49, H-1525 Budapest, Hungary}
\author{B\'alint Hartmann}
\affiliation{Centre for Energy Research, Institute for Energy Security\\
and Environmental Safety, P.~O.~Box 49, H-1525 Budapest, Hungary}
\author{Jeffrey Kelling}
\affiliation{Faculty of Natural Sciences, Technische Universit\"at
Chemnitz, \\ Stra{\ss}e der Nationen 62,  09111 Chemnitz, Germany}
\affiliation{Department of Information Services and Computing,
Helmholtz-Zentrum Dresden-Rossendorf, P.~O.~Box 51 01 19, 01314 Dresden, Germany
}

%%%%%%%%%%%%%%%%%%%%%%%%%%%%%%%%%%%%%%%%%%%%%%%%%%%%%%%%%%%%%%%%%%%%%%%%%
\begin{abstract}

Dynamical simulation of the cascade failures on the EU and USA high-voltage
power grids has been done via solving the second-order Kuramoto equation.
We show that synchronization transition happens by increasing the global
coupling parameter $K$ with metasatble states depending on the initial 
conditions so that hysteresis loops occur. 
We provide analytic results for the time dependence of frequency 
spread in the large $K$ approximation and by comparing it with numerics 
of $d=2,3$ lattices, we find agreement in the case of ordered initial 
conditions. However, different power-law (PL) tails occur, when the 
fluctuations are strong. 
After thermalizing the systems we allow a single line cut failure and 
follow the subsequent overloads with respect to threshold values $T$. 
The PDFs $p(N_f)$ of the cascade failures exhibit PL tails near
the synchronization transition point $K_c$. 
Near $K_c$ the exponents of the PL-s for the US power grid vary with 
$T$ as $1.4 \le \tau \le 2.1$, in agreement with the empirical blackout 
statistics, while on the EU power grid we find somewhat steeper PL-s
characterized by $1.4 \le \tau \le 2.4$. 
Below $K_c$ we find signatures of $T$-dependent PL-s, caused by frustrated 
synchronization, reminiscent of Griffiths effects.
Here we also observe stability growth following the blackout cascades, 
similar to intentional islanding, but for $K > K_c$ this does not happen.
For $T < T_c$, bumps appear in the PDFs with large mean values, 
known as ``dragon king'' blackout events.
We also analyze the delaying/stabilizing effects of instantaneous feedback 
or increased dissipation and show how local synchronization behaves on
geographic maps. 

\end{abstract}
%%%%%%%%%%%%%%%%%%%%%%%%%%%%%%%%%%%%%%%%%%%%%%%%%%%%%%%%%%%%%%%%%%%%%%%%%

\maketitle

%%%%%%%%%%%%%%%%%%%%%%%%%%%%%%%%%%%%%%%%%%%%%%%%%%%%%%%%%%%%%%%%%%%%%%%%%
\section{Introduction}
%%%%%%%%%%%%%%%%%%%%%%%%%%%%%%%%%%%%%%%%%%%%%%%%%%%%%%%%%%%%%%%%%%%%%%%%%

With the power sector undergoing huge changes, the modelling of power
grids has become a focus point for researchers in the fields of
statistical physics as well. The hierarchical modular structure (or HMN)
of these networks, their heterogeneity and their size make them a
perfect candidate for complex system analysis, as discussed in 
~\cite{Acebron,ARENAS200893,POWcikk}.
The shift from traditional fuels to large scale utilization of renewable
energy sources poses a number of challenges in regards of robustness and
resilience of power systems, mainly due to the appearing correlated
spatio-temporal fluctuations. Sudden changes (e.g. intermittent
production, load ramps, outages) may start desynchronization cascades,
which would propagate through the whole synchronously operated system as
an avalanche. The resulting blackouts are of various sizes, but they
often lead to full system desynchronizations lasting for extended
periods~\cite{1}. The size distributions of the outages have been found
scale-free in the US, China, Norway, Sweden in the available long time
series data. In particular, power-laws could be fitted in countries,
with so called robust networks~\cite{CasTh}, where the networks are 
categorized as robust or fragile based on static network topology analysis. 
As defined by~\cite{Pgridtop}, networks with $P(k > K) = C \exp{(-k/\gamma)}$ 
cumulative degree ($k$) distribution and $\gamma < 3/2$ are called robust.

The research community is actively working on methods to support
understanding and forecasting of such events~\cite{Rev-Eng} 
More particularly, taking the aspect of statistical physics, risk of 
failure in power systems represents a specific case of the risk of 
system-wide breakdown in threshold activated disordered systems. 
Self-organized criticality (SOC)~\cite{SOC} is typically used for 
the modelling of such phenomena~\cite{Car}, where SOC is expected to 
arise as the consequence of self-tuning to a critical point, which 
is determined by the competition of power needs and transmission 
capabilities of the grid itself. DC threshold models~\cite{Car2} 
can be extended by considering AC power flows of power systems,
modelling them via the second order Kuramoto equation~\cite{fila}. 
An increasing number of papers discuss synchronization and stability 
issues by this approach, such as in
Refs.~\cite{Carareto,CHOI201132,5,12,13,18,29,32,36,Gryz,Olmi-Sch-19,pol}.

These solutions can be deduced from the power transmission behavior 
of AC systems and are actually a generalization of the Kuramoto model
~\cite{Kura} with inertia. Below $d < d_l = 4$ the Kuramoto model 
does not exhibit real phase transitions to a synchronized state, but 
a smooth crossover~\cite{chate_prl}. In real life, partially
synchronized states can be observed. It was also observed that in lower
graph dimensions the transition point shifts towards infinity as the
system size and hysteresis behavior emerge~\cite{POWcikk}.

Highly heterogeneous systems (often referred to as disordered with
respect to the homogeneous ones) can exhibit rare region effects, which
alter critical dynamics~\cite{Vojta}. Such rare regions, locally in 
different state than the remainder of the network, evolve slowly 
and contribute to the global order parameter, causing slow dynamics 
and fluctuations. They can generate so-called Griffiths Phases 
(GP)~\cite{Griffiths} in an extended region around the critical point, 
causing slowly decaying auto-correlations and burstiness~\cite{burstcikk}.
Their existence in load driven power-grid and earthquake models is 
studied very recently by~\cite{Bis}.
These rare regions can also lead to frustrated synchronization and 
chimera states~\cite{Frus,Frus-noise,FrusB}, which result
in non-universal PL distributions of the desynchronization events below
the transition point~\cite{POWcikk,KurCC,KKI18deco,Flycikk}. 
The authors have previously provided evidence for this using the 
second order Kuramoto model on 2d lattices
and large synthetic power system topologies~\cite{POWcikk} and on the 
realistic representation of the Hungarian power system while also 
allowing line outages (line-cuts)~\cite{Powfailcikk}.

The second order Kuramoto model with power transmission thresholds (line
capacities) has been also introduced in~\cite{SWTL18,Powfailcikk} 
for the dynamical modelling of cascade events. 
Identification of critical lines of transmission in
different national power grids has been determined. We follow this
method in order to investigate the desynchronization duration
distributions via measuring the number of failed lines following a link
removal event. We shall also compare results obtained on 2d and 3d
lattices with those of the US end EU high voltage power-grid. Multiple
recent studies suggested that it is possible to prevent the spread of
cascade failures either by adding an isolator as a fortress against an
attack~\cite{kaiser2021net}, or by adding some nodes to increase the 
rerouting path, or by strengthening a link~\cite{kaiser2021top}. 
Surprisingly, we observed that it is also possible to increase the 
synchronization level by triggering moderate cascade line failures for 
more weakly coupled power grids. This effect
bears a close resemblance to the so-called ``intentional islanding
operation'' in which the spread of failures could be contained by
removing some weak links so that a power grid is segregated into several
self-sustained islands~\cite{baldick2008ini,esmaeilian2016}.
As coupled oscillators described by the second order Kuramoto model 
exhibit a discontinuous synchronization
transition, cascade distributions are expected to consist of two
different types of avalanches, regular ones and huge avalanches or kings
~\cite{Sorn-Uui}, corresponding to the bumps in the distributions. 
Recently these kings, also called ``Dragon Kings'' (DK), were shown 
to exist in sand-pile models, following self-organized bistability
~\cite{SOB}, which is expected to be a common phenomenon in nature. 
Very recently, in a sand-pile model, coupled to massless Kuramoto oscillators, 
designed to model power-grids the existence of DK-s has also been shown 
~\cite{Gur-Sou}. We also show the appearance of DK-s in our power-grid 
simulations, based on the swing equations of the AC circuits.

The remainder of this paper is organized as follows: In
Sec.~\ref{sec:2} we describe the methods and the power grids to be used
in the paper. In Sec.~\ref{sec:3}, we first explore the temporal
evolution of the frequency spread and the phase order parameter during
thermalization and after invoking one random line cut. We also give a
qualitative explanation for the frequency spread algebraic decay via a
linear approximation argument~\cite{HPCE}. We then present numerical
results for cascades for the EU-HV and US-HV networks. These results
show that in the neighborhood of criticality, cascade sizes can display
non-universal power laws and there is also a manifestation of islanding
effects~\cite{baldick2008ini,esmaeilian2016}. In Sec.~\ref{sec:4}, we
briefly demonstrate that the local Kuramoto order parameter is not
uniform across different geographical regions. Finally, Sec.~\ref{sec:5}
summarizes this work.

%%%%%%%%%%%%%%%%%%%%%%%%%%%%%%%%%%%%%%%%%%%%%%%%%%%%%%%%%%%%%%%%%%%%%%%%%
\section{Models and methods \label{sec:2}}
%%%%%%%%%%%%%%%%%%%%%%%%%%%%%%%%%%%%%%%%%%%%%%%%%%%%%%%%%%%%%%%%%%%%%%%%%

In the lack of full details of power-grids, as a first approximation,
the evolution of synchronization is described by the swing 
equations~\cite{swing} set up for mechanical elements 
with inertia by the second-order Kuramoto equation~\cite{fila}.
For a network of $N$ oscillators with phase $\theta_i(t)$:
\begin{eqnarray}\label{kur2eq}
\dot{\theta_i}(t) & = & \omega_i(t) \\
\dot{\omega_i}(t) & = & \omega_i(0) - \alpha \dot{\theta_i}(t) 
+ K \sum_{j=1}^{N} A_{ij} \sin[ \theta_j(t)- \theta_i(t)] \ , \nonumber
\end{eqnarray}
where $\alpha$ is the damping parameter, describing the power dissipation,
or an instantaneous feedback ~\cite{Powfailcikk}, $K$ is a global coupling, 
related to the maximum transmitted power between nodes and $A_{ij}$, 
which is the adjacency matrix of the network containing admittance elements. 
The self-frequency of the $i$-th oscillator $\omega_i(0)$ describes the 
power in/out of a given node when Eq.~\eqref{kur2eq} is considered to 
be the swing equation of a coupled AC circuit.

In our simulations the following parameter settings were used:
the $\alpha$ dissipation factor, which is chosen to be equal to 
$0.4$ to meet expectations for power-grids, with the $[1/s]$
inverse time physical dimension assumption.
For modeling instantaneous feedback, or increased damping 
parameter we applied: $\alpha = 3.0 [1/s]$ similarly as 
in~\cite{Powfailcikk}.

To solve the differential equations we used the adaptive Bulirsch-Stoer
stepper~\cite{boostOdeInt} in general, which provides more precise 
results for large $K$ coupling values.
To obtain reasonable statistics, via the adaptive solver we
needed very strong computing resources, using parallel codes running on
GPU clusters by utilizing the VexCL library for vector operations 
\cite{vexcl}, but the results were cross-checked with the results
obtained on conventional CPU machines and compared with the 
Runge-Kutta-4 solvers too. The details of the GPU implementation
will be discussed in a separate publication.

Exploiting the Galilean invariance of Eq.~(\ref{kur2eq}) we can gauge out 
the mean value $\langle \omega_i(0)\rangle$ in a rotating frame, 
thus for the self-frequencies $\omega_i(0)$ we used Gaussian random 
variables with zero mean and unit variance.

What is more, another rescaling invariance can also be discovered, 
similarly as in the case of the first-order Kuramoto 
model~\cite{KKI18deco}:
\begin{equation}\label{rescale}
	K'=a K, \ \omega'_i(0) = a \omega_i(0), \ \alpha' = \sqrt{a} \alpha, 
	\ t' =t / \sqrt{a}\,,
\end{equation}
which could also be exploited by the solutions at very large
couplings, where the adaptive solver may slow down
substantially. Conversely, if the relaxation time (for smaller $K$) 
is too slow we could avoid it by increasing $K$. 
However, there is an optimal value for rescaling due to the balance
between the speed gain for the solver by using a smaller $K$ and the
dilation of time by a factor of $\sqrt{a}$. This may be determined 
by actual simulations for a given network or parameter set. 
We measured that by choosing $a=1/9$, with which, although the 
simulation time $t'$ is increased by a factor of $3$,
the GPU run times became faster by a factor of $20\%$ as compared to 
the runs without rescaling in case of large 2d lattices with 
linear sizes $L=3000$. Further benchmark results will be 
published in a separate publication~\cite{texpaper}.

To achieve thermalized states with larger synchronization the
initial state was set to be fully synchronized, with phases: 
$\theta_i(0)=0$, but to determine the hysteresis curve or 
to investigate $\Omega(t)$ decays we also used the uniform 
random distribution: $\theta_i(0) \in (0,2\pi)$. 
The initial frequencies were set as $\dot{\theta_i}(0)=\omega_i(0)$.
The thermalization was performed by running the code for 
$1000 - 5\times 10^4$ time steps, judging by visual or 
automatic inspection of the status of order parameter saturation, 
without allowing line cuts on the graph.
This, is the so called ''quench'' procedure in statistical physics, which 
does not happen in reality, because in case of a cold restart both the 
consumers and generators are connected gradually and adiabatically, 
following a prescribed protocol to avoid initial unbalances. 
However, instead of the more tedious equilibration process we 
quenched the system from random or ordered initial conditions to a 
steady states without allowing cascade failures. 
This also has the advantage that we could learn the dynamics of the 
second-order Kuramoto model on different networks.

Following the thermalization we perturbed the system by removing a 
randomly selected link, or alternatively, a randomly selected node, 
in order to simulate a power failure event. 
Following that, if the ensuing power flow on a line between 
neighboring nodes was greater than a threshold:
\begin{equation}\label{Fij}
	F_{ij} = | \sin(\theta_j-\theta_i) | > T\,,
\end{equation}
so that that line is regarded as overloaded,
we removed this link from the graph permanently and measured the
total number of line failures $N_f$ of the simulated blackout 
cascades of each realizations, corresponding to different 
$\omega_i(0)$ self frequency values.
At the end we applied histogramming to determine the PDFs of $N_f$. 
In the vicinity of criticality, one usually expects power-law 
distributions of the form
\begin{equation}
	p(N_f)\sim N_f^{-\tau}\, ,
	\label{Nfpdf}
\end{equation}
thus we plotted our results on the log-log scale.

During the cascade simulations, which had the length of 
$t_{max}=10^3$ -- $10^4$ we measured the Kuramoto phase order 
parameter:
\begin{equation}\label{ordp}
	z(t_k) = r(t_k) \exp{\left[i \theta(t_k)\right]} = \frac{1}{N} \left| \sum_j
\exp{\left[i \theta_j(t_k)\right]}\right| \ ,
\end{equation}
by increasing the sampling time steps exponentially :
\begin{equation}
	t_k=t_0\times 1.08^k \ .
\end{equation}
In Eq.~\eqref{ordp}, $0 \le r(t_k) \le 1$ gauges the overall coherence 
and $\theta(t_k)$ is the average phase. 
The set of equations (\ref{kur2eq}) were solved numerically for 
$10^3 - 3\times 10^4$ independent initial conditions, 
initialized by different $\omega_i(0)$-s, and different
$\theta_i(0)$-s if applicable, and sample averages for the 
phase order parameter
\begin{equation}\label{KOP}
R(t_k) = \langle r(t_k)\rangle
\end{equation} 
and for the variance of the frequencies
\begin{equation}\label{FOP}
\Omega(t,N) = \av{\frac{1}{N} \sum_{j=1}^N
(\overline\omega-\omega_j)^2}
\end{equation}
were determined.
In case of a single peaked self-frequency distribution, $\Omega(t,N)$ 
is an appropriate order parameter and is easier to measure, besides 
the more commonly used measure~\cite{HPCE}, which counts the number 
of oscillators in the largest cluster having an identical frequency.

In the steady state after thermalization we also measured the standard 
deviation of the order parameters $R(t_k)$ and $\Omega(t,N)$ in 
order to locate the transition point $K_c$ by the fluctuation maxima. 
In case of the first-order Kuramoto equation the fluctuations of
both order parameters show a maximum at the same $K_c$~\cite{Flycikk},
meaning a maximal chaoticity here.
For the second-order Kuramoto equation only $\sigma(R(t_k))$ seems 
to have a peak at $K_c$; see Sec.~\ref{sec:3C} and Sec.~\ref{sec:3D} for
more details. Nonetheless, one would expect chaotic dynamics to
emerge in the vicinity of the transition point.

%%%%%%%%%%%%%%%%%%%%%%%%%%%%%%%%%%%%%%%%%%%%%%%%%%%%%%%%%%%%%%%%%%%%%%%%%
\subsection{Description and analysis of the power grids \label{sec:2A}}
%%%%%%%%%%%%%%%%%%%%%%%%%%%%%%%%%%%%%%%%%%%%%%%%%%%%%%%%%%%%%%%%%%%%%%%%%

Here we determined some basic topological characteristics~\cite{Newmanbook} 
of the studied power grids using the Gephi tool~\cite{gephi} and 
MATHEMATICA. For comparison, we
have studied the dynamical behavior of the western US-HV power grid,
obtained from~\cite{USpg}, and of the EU-HV power grid, obtained from
the ``SciGRID Dataset''~\cite{EUgrid}. These power-grid networks are
genuinely hierarchical modular networks if the detailed information
for the medium- and low-voltage parts of the grids are also incorporated. 
Practically, it is almost impossible to infer the entire structure of a 
large power-grid network, but it is feasible to mimic a realistic power 
grid network by adding to the HV skeleton of the grid the medium- and
low-voltage parts according to the empirical hierarchical
distribution, as we previously did in Ref.~\cite{POWcikk} for the
Hungarian power-grid network. In this work, we will only focus on the
high voltage networks as obtained from the raw data. For
simplicity, all transmission lines are regarded as bidirectional and 
identical. Nodes are also identical and featureless.

\begin{figure}[h]
\centering
\includegraphics[width=6.5cm]{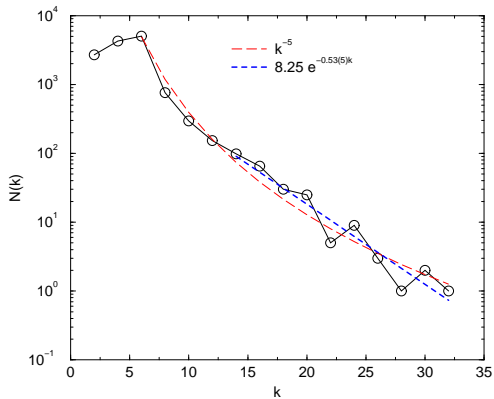}
\caption{Degree distribution of the EU-HV network. The blue dashed
	line shows an exponential fit for $k\ge 15$,
        while the red long-dashed line is a a PL fit for $k > 5$.}
\label{EUkdeg}
\end{figure}

Standard graph measures for the US-HV and the EU-HV networks are 
summarized in Table~\ref{UStab}. The US-HV network statistics were also
summarized in~\cite{Powfailcikk}. The US-HV network consists of $N=4194$ 
nodes and $E=6594$ edges, while $N=13478$ nodes of the EU-HV network are 
interconnected via $E=33844$ links. The average degrees $\av{k}$ of the 
two networks take very similar values $2.67$ and $2.51$, respectively. 
As an example, in Fig.~\ref{EUkdeg}, we show the degree distribution
of the EU-HV network. The tail of the degree distribution ($k > 5$) may
be fitted with a PL with a large exponent $\simeq 5$. However, it is more 
reasonable to fit the data by an exponential function, which is common for 
power grids~\cite{ROSASCASALS2011805,BalintSRcikk}. For $k\ge 15$ a 
stretched exponential $8.25 \times e^{-0.53(5)k}$ fits the data
quite well. Furthermore, comparing the exponents $\gamma$ of the cumulative 
degree distribution of the two networks renders EU-HV network to be just 
at the threshold of robust/fragile: $\gamma=3/2$, according to the 
definition by~\cite{Pgridtop}, while the US-HV network can be categorized 
as a robust network.
\begin{table}
 \caption{\label{UStab} Graph measures of the investigated
 power-grids.}
  \scriptsize
 \centering
 \begin{tabular}{c|cccccccccc}
  \hline\hline
  & $N$  & $E$  & $\av{k}$ & $\gamma$ & $Q$ & $L$ & $L_r$ & $C$ &
    $\sigma$ & $d$ \\
  \hline
  US  &4194 & 6594 & 2.67  & 1.24(1) & 0.929 &  18.7 & 17.7 & 0.08  & 9.334 & 3.0(1) \\
  \hline
  EU  & 13478 &33844& 2.51 & 1.53(5) & 0.963  & 49.505 & 10.2 &  0.089 &   98.63 & 2.6(1) \\
 \hline\hline
 \end{tabular}
\end{table}

%%%%%%%%%%%%%%%%%%%%%%%%%%%%%%%%%%%%%%%%%%%%%%%%%%%%%%%%%%%%%%%%%%%%%%%%%
%%%%%%%%%%%%%%%%%%%%%%%%%%%%%%%%%%%%%%%%%%%%%%%%%%%%%%%%%%%%%%%%%%%%%%%%%

%%%%%%%%%%%%%%%%%%%%%%%%%%%%%%%%%%%%%%%%%%%%%%%%%%%%%%%%%%%%%%%%%%%%%%%%%
\begin{figure}[h]
\centering
\includegraphics[width=6.5cm]{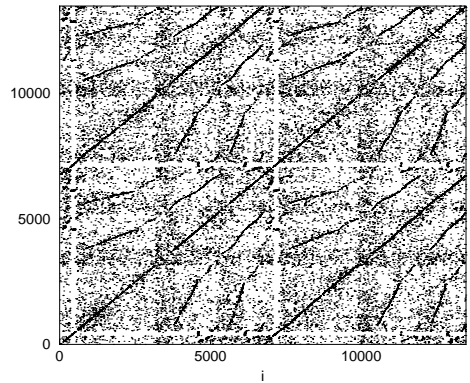}
\caption{The adjacency matrix of the EU-HV grid, with $A_{ij}=1$ 
displayed in black and white if otherwise (c.f.~\cite{Powfailcikk}
for the US-HV grid data).}
\label{EU-HV-A}
\end{figure}
%%%%%%%%%%%%%%%%%%%%%%%%%%%%%%%%%%%%%%%%%%%%%%%%%%%%%%%%%%%%%%%%%%%%%%%%%

The adjacency matrix of the EU-HV network is shown in
Fig.~\ref{EU-HV-A}. Similar to the US-HV network, this is a highly 
modular network with modularity quotient $Q=0.964$, defined by
\begin{equation}
Q=\frac{1}{N\av{k}}\sum\limits_{ij}\left(A_{ij}-
\frac{k_ik_j}{N\av{k}}\right)\delta(k_i,k_j),
\end{equation}
where $\delta(i,j)$ is the Kronecker delta function.  

The Watts-Strogatz clustering coefficient \cite{WS98} of a
network of $N$ nodes is
\begin{equation}\label{Cws}
C = \frac1N \sum_i 2n_i / k_i(k_i-1) \ ,
\end{equation}
where $n_i$ denotes the number of directed edges, interconnecting the
$k_i$ nearest neighbors of node $i$. With $C_{US}=0.08$ and 
$C_{EU}=0.089$, both networks show a much higher clustering coefficient 
than that of a random graph with the same $N$ and $E$, defined by 
$C_r = \langle k\rangle / N$. The $C_{EU} = 0.089$ is about $480$ times 
higher, than that of a random network of same size $C_r=0.000186229$.

The average shortest path length is
\begin{equation}
L = \frac{1}{N (N-1)} \sum_{j\ne i} d(i,j) \ ,
\end{equation}
where $d(i,j)$ is the graph distance between vertices $i$ and $j$.
For a random graph of the same size and edge number, Ref.~\cite{Fron}
gives the expression
\begin{equation}
L_r = \frac{\ln N - 0.5772}{\ln\langle k\rangle} + 1/2 \ .
\end{equation}
As shown in Table~\ref{UStab}, in case of the EU-HV network, $L$
is much larger than $L_r$, in stark contrast to the US-HV network.
Furthermore, according to the definition of the
coefficient~\cite{HumphriesGurney08}:
\begin{equation}
\sigma = \frac{C/C_r}{L/L_r} \ ,
\label{swcoef}
\end{equation}
the EU-HV network is a small-world network because $\sigma \simeq 100 $, is 
much larger than unity.

To summarize, from Table~\ref{UStab} we can see that the EU-HV 
network has about $3$ times more nodes and $5$ times more edges than the 
western US-HV grid and exhibits similar measures, except for 
$L$, which is also $\sim 2.5$ times bigger. Furthermore, 
by the degree exponent it is marginally fragile, unlike the US-HV grid
which is robust, thus the existence of PL blackout size distributions
is a matter of question. 
We shall investigate if this holds in the dynamical sense, 
in the presence of fluctuating energy resources. 

\begin{figure}[ht]
\centering
\includegraphics[width=6.5cm]{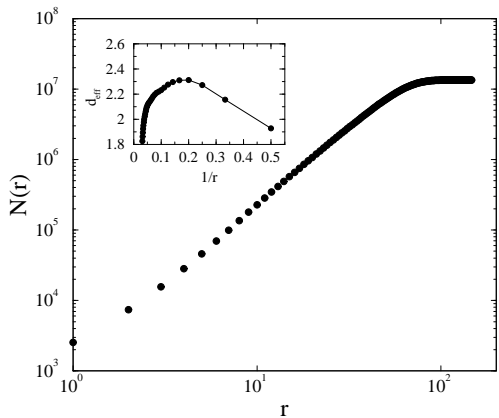}
\caption{Number of nodes with distance $r$ or less from random seeds
in the breadth-first search algorithm applied for the EU-HV graph.  
The inset shows the local slopes, corresponding to the same data.
Linear extrapolation to $1/r\to 0$ provides $d=2.6(1)$.}
\label{EUdim}
\end{figure}

In addition, we also provide estimates for the effective graph 
(topological) dimension $d$, defined by
\begin{equation} \label{topD}
N(r) \sim r^d,
\end{equation} 
where we counted the number of nodes $N(r)$ with chemical distance $r$ or 
less from randomly selected seeds and calculated averages over many trials. 
In Fig.~\ref{EUdim}, we illustrate the growth of $N(r)$ for the EU-HV 
network. Note, however that finite-size cutoff happens already for 
$r > 30$. 
To see the corrections to scaling we determined the effective exponents
of $d$ as the discretized, logarithmic derivative of \eqref{topD}
\begin{equation}  \label{Deff}
	d_\mathrm{eff}(r+1/2) = \frac {\ln \langle N(r)\rangle - \ln \langle N(r+1)\rangle} {\ln(r) - \ln(r+1)} \ .
\end{equation}
These local slopes are shown in the inset of Fig.~\ref{EUdim},
as the function of $1/r$ and enables an extrapolation to 
$1/r\to 0$. This gives a smaller value $d=2.6(1)$ for the EU-HV network, 
than in case of the US-HV network $d=3.0(1)$.

%%%%%%%%%%%%%%%%%%%%%%%%%%%%%%%%%%%%%%%%%%%%%%%%%%%%%%%%%%%%%%%%%%%%%%%%%
%%%%%%%%%%%%%%%%%%%%%%%%%%%%%%%%%%%%%%%%%%%%%%%%%%%%%%%%%%%%%%%%%%%%%%%%

%%%%%%%%%%%%%%%%%%%%%%%%%%%%%%%%%%%%%%%%%%%%%%%%%%%%%%%%%%%%%%%%%%%%%%%%%
\section{Synchronization dynamics on the power grids \label{sec:3}}
%%%%%%%%%%%%%%%%%%%%%%%%%%%%%%%%%%%%%%%%%%%%%%%%%%%%%%%%%%%%%%%%%%%%%%%%%

%%%%%%%%%%%%%%%%%%%%%%%%%%%%%%%%%%%%%%%%%%%%%%%%%%%%%%%%%%%%%%%%%%%%%%%%%
\subsection{Frequency entrainment for large $K$ \label{sec:3A}}
%%%%%%%%%%%%%%%%%%%%%%%%%%%%%%%%%%%%%%%%%%%%%%%%%%%%%%%%%%%%%%%%%%%%%%%%

\begin{figure}[!htbp]
	\centering
	\includegraphics[width=0.49\textwidth]{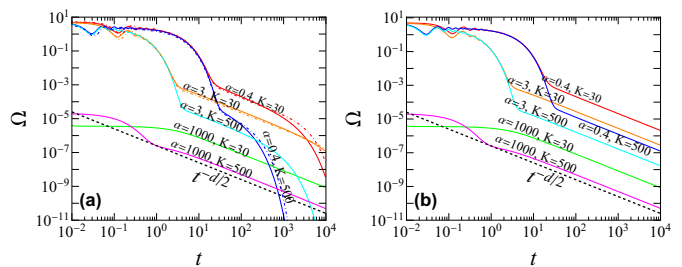}
	\caption{The frequency spread for $d=2$, obtained through 
	numerical
        integration of Eq.~\eqref{spreadlin} at different $\alpha$ and 
	$K$ values for (a) $L=3000$ and (b) $L=10^{12}$. For 
	comparison, the dot dashed lines show the full results of 
	Eq.~\eqref{kur2eq} with 
	synchronized initial conditions and the dashed lines mark the 
        algebraic decay law $t^{-d/2}$ of the first-order Kuramoto model.} 
	\label{fig:lspread}
\end{figure}

It is known that on a $d$-dimensional lattice, the frequency order 
parameter (\ref{FOP}) decays as 
$\Omega \propto t^{-d/2}$ in case of the first-order Kuramoto
model in the large coupling limit \cite{HPCE}.
Following Ref.~\cite{HPCE}, we have investigated this in the case of 
the second-order Kuramoto model, first by applying a linear approximation 
by replacing $\sin(x) \propto x$ to gain a qualitative understanding. 
In a $d$-dimensional space, casting the continuum second-order Kuramoto 
equations into the momentum space, we are led to solve
\begin{equation}
	\frac{\partial^2}{\partial t^2}\theta(\mathbf{k},t)+
	\alpha \frac{\partial }{\partial t} \theta(\mathbf{k},t) = 
	\omega(\mathbf{k},0)-Kk^2 \theta(\mathbf{k},t)\,.
	\label{linearapp}
\end{equation}
Without citing the cumbersome solution, we simply give the expression for
the phase velocity [$\omega(\mathbf{x},t)\equiv\dot{\theta}
(\mathbf{x},t)$] in the Fourier space
\begin{align}
	\omega(\mathbf{k},t)=&e^{-\frac{1}{2} t (\alpha +\Delta )} 
	\Big[\omega(\mathbf{k},0) \big(( \Delta +2 -\alpha) e^{\Delta  t}
	\nonumber\\
	& +\alpha +\Delta -2\big)-2 K k^2 
	\theta(\mathbf{k},0) \left(e^{\Delta  t} -1\right)\Big]/2\Delta\,,
	\label{velocity}
\end{align}
where $\Delta=\sqrt{\alpha^2-4K k^2}$. In the initial short-time regime,
$\omega(\mathbf{k},t)$ displays a transient oscillation 
(also cf.~Fig.~\ref{fig:lspread}) for modes $k>\alpha/2\sqrt{K}$, but 
they are quickly suppressed by the factor $e^{-\frac{1}{2}t}$. Similar 
to the first-order Kuramoto model \cite{HPCE}, all Fourier modes of 
$\omega(\mathbf{k},t)$ and now \textit{including} the $\mathbf{k}=0$ 
mode also vanish in the long-time limit, suggesting that the phase 
velocity becomes uniform and the frequency spread $\Omega$ approaches 
zero. This is immediately verified by the explicit expression for 
$\Omega$:
\begin{align}
\Omega(t)=&\frac{1}{L^d}\int d^d\mathbf{x} \langle 
\left[\omega(\mathbf{x},t)-\bar{\omega}(t)\right]^2\rangle 
	   \nonumber\\
	 =&C_d\int_{2\pi/L}^{\pi/a}dk k^{d-1}\frac{e^{-t (\alpha 
	 +\Delta )}}{4 \Delta ^2}
	 \Big[\alpha +\Delta -2 \nonumber\\
	 &+(\Delta-\alpha +2) e^{\Delta  t}\Big]^2\,,
	\label{spreadlin}
\end{align}
where $\bar{\omega}(t)$ denotes the spatial average of 
$\omega(\mathbf{x},t)$, while $a$ and $C_d$ are the lattice spacing and 
the geometric factor, respectively, both of which can be innocuously 
taken as 1 for a qualitative analysis.
%%%%%%%%%%%%%%%%%%%%%%%%%%%%%%%%%%%%%%%%%%%%%%%%%%%%%%%%%%%%%%%%%%%%%%%%
%%%%%%%%%%%%%%%%%%%%%%%%%%%%%%%%%%%%%%%%%%%%%%%%%%%%%%%%%%%%%%%%%%%%%%%%

As shown in Fig.~\ref{fig:lspread} (b), for a large enough system,
after a short initial oscillatory transition, the frequency spread 
display the same algebraic decay (in the intermediate short-time regime
$1\ll t\ll L/\sqrt{K}$) as that of the first-order Kuramoto 
model, as long as $\alpha>0$ and $K>0$. This is most evident for the
scenario when $\alpha\gg 1$ as the damping term becomes essentially 
dominant over the second-order term. In smaller systems, due to 
finite-size effects, the algebraic decay may give way to a more rapid 
decay for smaller $\alpha$ and larger $K$ values 
[cf.~Fig.~\ref{fig:lspread}(a)], with which, one may still observe a 
distorted algebraic decay in the intermediate time range.

It should be emphasized that linear approximation inevitably missed 
out nonlinearity of the system as well as the effects of heterogeneity 
in the initial frequency/phase distribution and in the topological 
structure. 
Hence, an intermediate algebraic decay may also be observed for the 
frequency spread by explicitly solving Eq.~\eqref{kur2eq}, but it doesn't 
necessarily follow $t^{-d/2}$, e.g.~due to nonlinearity. 
In particular, most realistic complex networks are characterized by a
relatively small average shortest path length (see 
e.g.~Table~\ref{UStab}) as well as a more heterogeneous structure, the
latter of which could even render linear approximation invalid. Both
these two factors may then strongly suppress any manifestation of an 
elongated algebraic decay for networks, except for a certain critical
coupling value, which is exactly what we observed for the EU-HV and 
the US-HV networks, as will be shown in below with respect
to the US-HV network. 

\begin{figure}[ht]
\centering
\includegraphics[width=6.5cm]{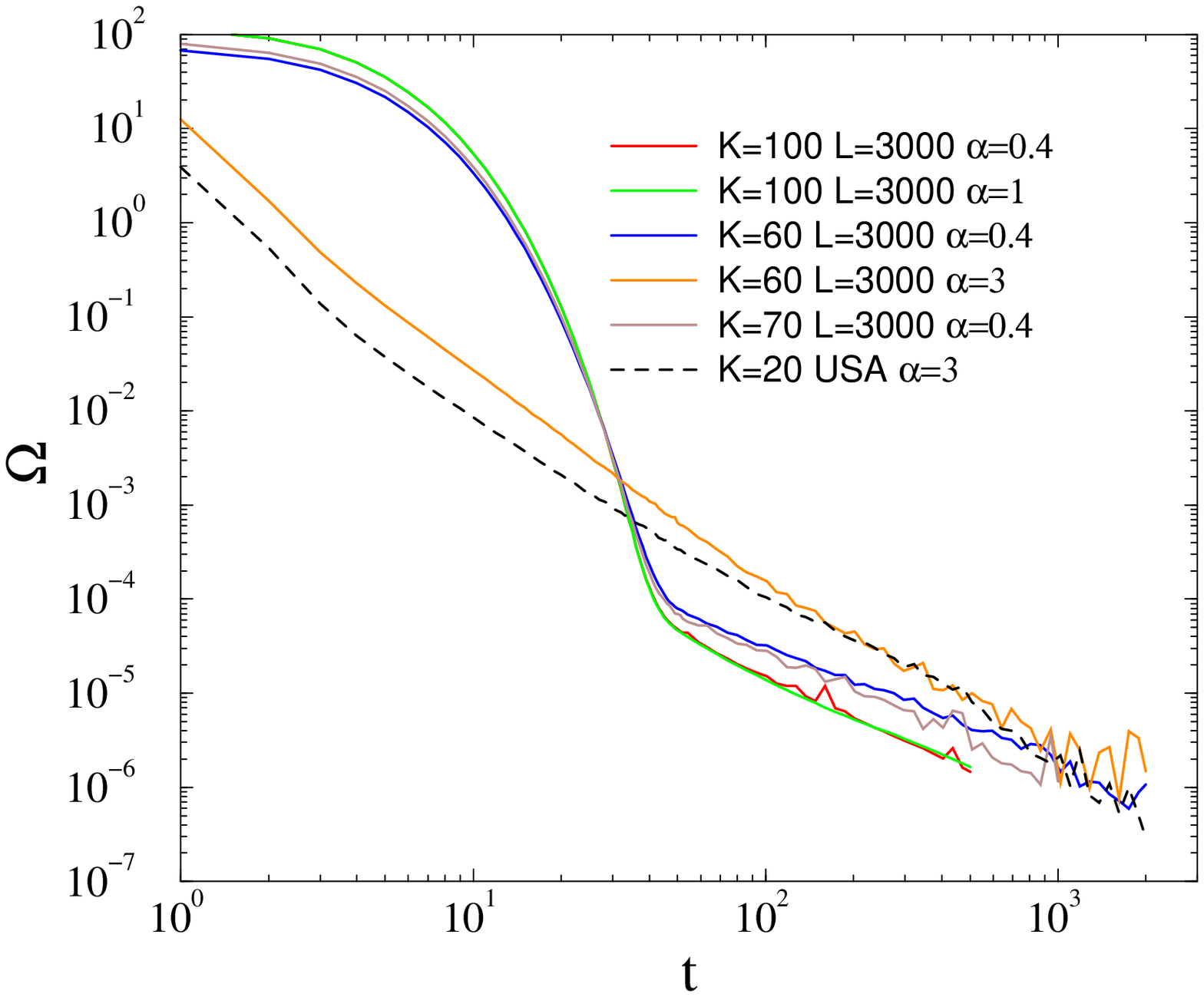}
\caption{\label{O2Dtherm} Evolution of frequency spread order parameter on 
the 2d lattices as the function of $K$ and $\alpha$ values shown 
by the legends in case of disordered initial conditions. 
These curves exhibit PL tails within the exponent range $1.2 \le \tau \le 1.5$. 
For comparison the dashed line shows the result for the US-HV network at 
$K=20$ and $\alpha=3$, for which $\tau\simeq 1.5$ can be read off.}
\end{figure}

\begin{figure}[ht]
\centering
\includegraphics[width=6.5cm]{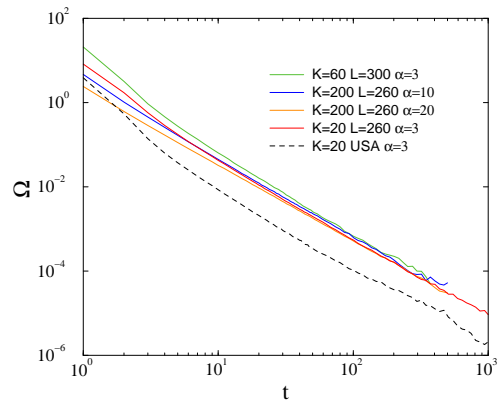}
\caption{\label{O3Dtherm} Evolution of frequency spread order parameter on 
the 3d lattices as the function of $K$ and $\alpha$ as shown 
by the legends in case of disordered initial conditions.
These curves exhibit PL tails, with similar exponents $\tau \simeq 1.85$.
For comparison the dashed line shows the result for the US-HV at $K=20$
and $\alpha=3$, where $\tau\simeq 1.5$ can be read off.}
\end{figure}

To this end, we solved Eq.~\eqref{kur2eq} on $L=3000$ (2d) and 
$L=260$ (3d) linear sized lattices, with periodic boundary conditions, 
as well as on the EU-HV and the US-HV networks. Averaging was done over
$100$ -- $1000$ samples of different initial self frequencies.
Although the linear approximation shows good agreement with simulation
results, when we start from a fully synchronized state, as 
Figs.~\ref{O2Dtherm} and \ref{O3Dtherm} show, the $\Omega(t)$
spread decays exhibit non-trivial $K$- and $\alpha$-dependent PL-s
for large global couplings if a disordered initial state
$[\theta_i(0)\in (0,2\pi)]$ is implemented. The fitted decay exponents 
$\tau$ change in the range $1.2 \le \tau \le 1.5$ in the 2d case.
Solutions on the 3d lattice provide PL tails, with exponents 
$\tau\simeq 1.85$. Therefore, the decay exponents are slightly 
larger than $d/2$, due to the non-negligible phase fluctuations.

In contrast to lattices, in the case of the US-HV network
on which the linear approximation fails, algebraic decay is only 
observed at $K_c=20$ with a larger $\alpha=3$. The PL tails, before size cutoff,
seem to agree with the regular lattice results, the fitted exponent 
is around $ \tau \simeq 1.5 \approx d/2$. 
Note that the effective graph dimension of the US-HV network is 
between 2 and 3, but an extrapolation to $N\to\infty$ gives 
$d=3.0(1)$ shown in Table~\ref{UStab}.

%%%%%%%%%%%%%%%%%%%%%%%%%%%%%%%%%%%%%%%%%%%%%%%%%%%%%%%%%%%%%%%%%%%%%%%%

%%%%%%%%%%%%%%%%%%%%%%%%%%%%%%%%%%%%%%%%%%%%%%%%%%%%%%%%%%%%%%%%%%%%%%%%

%%%%%%%%%%%%%%%%%%%%%%%%%%%%%%%%%%%%%%%%%%%%%%%%%%%%%%%%%%%%%%%%%%%%%%%
%%%%%%%%%%%%%%%%%%%%%%%%%%%%%%%%%%%%%%%%%%%%%%%%%%%%%%%%%%%%%%%%%%%%%%%%

%%%%%%%%%%%%%%%%%%%%%%%%%%%%%%%%%%%%%%%%%%%%%%%%%%%%%%%%%%%%%%%%%%%%%%%%%
\subsection{The effects of line cut and instantaneous
feedback\label{sec:3B}}
%%%%%%%%%%%%%%%%%%%%%%%%%%%%%%%%%%%%%%%%%%%%%%%%%%%%%%%%%%%%%%%%%%%%%%%%

\begin{figure}[ht]
\centering
\includegraphics[width=0.48\textwidth]{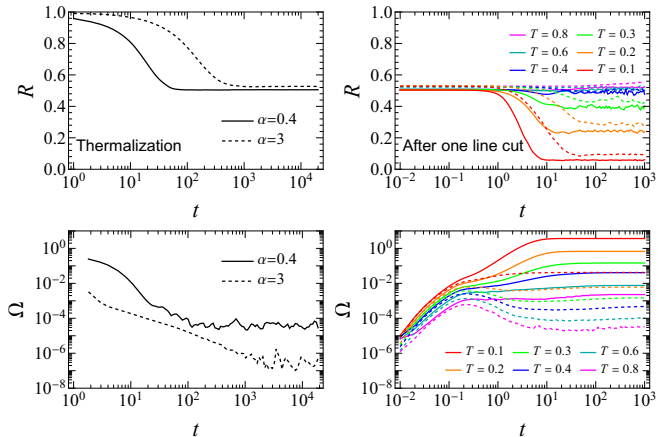}
\caption{\label{USthermcut} Evolution of phase order parameter (top) and
frequency spread order parameter (bottom) of the US network at $K=20$ 
during thermalization (left) and after one subsequent line cut (right) 
with respect to different thresholds (right: $T=0.1, 0.2,\dots,0.8$). 
Comparing to the normal case $\alpha=0.4$ (solid lines), a negative 
feedback $\alpha=3$ (dashed lines) essentially slows down the dynamics 
of the phase order parameter, while at the same time leads to a smaller 
frequency spread; similar effects are observed following one line cut.
}
\end{figure}

\begin{figure}[ht]
\centering
\includegraphics[width=0.48\textwidth]{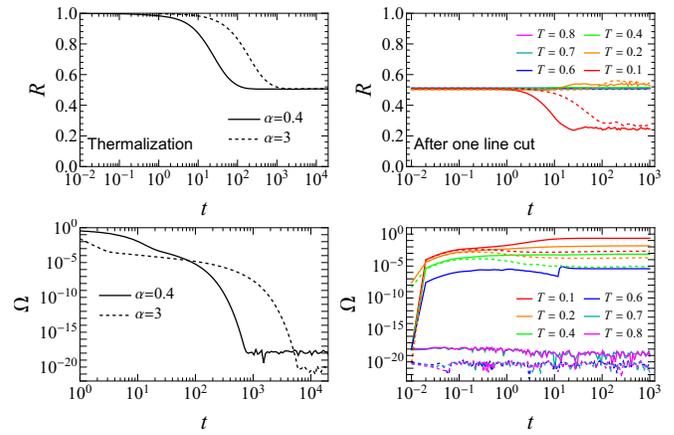}
\caption{\label{EUthermcut} Evolution of phase order parameter (top) and
frequency spread order parameter (bottom) of the EU network at $K=80$ 
during thermalization (left) and after one subsequent line cut (right) 
with respect to different thresholds (right: $T=0.1,0.2,0.4,0.6,0.8$). 
Comparing to the normal case $\alpha=0.4$ (solid lines), a negative 
feedback $\alpha=3$ (dashed lines) essentially slows down the dynamics 
of the phase order parameter, while at the same time leads to a smaller 
frequency spread; similar effects are observed following one line cut.
}
\end{figure}

In this section, we introduce a random transmission line cut to
thermalized, stable systems. As shown in Figs.~\ref{USthermcut} 
and \ref{EUthermcut}, after the systems transited into their stationary 
states with a thermalization process without allowing any line
failures, a subsequent random line cut introduces a perturbation, 
that can trigger a cascade of line failures dictated by the overload 
condition \eqref{Fij} and hence decrease the synchronization
level. In accord with our intuition, systems with a lower threshold value 
(i.e.~lines are more vulnerable against disturbances) usually are affected
to a greater extent by one line cut, transiting into a new stationary 
state with a lower phase order parameter value and a broader frequency
spread. However, this may not always be the case. As can be seen in the case 
of the EU network, the phase order parameter may even increase after one line
cut for certain intermediate threshold values (e.g.~$T=0.2$), irrespective
of the fact that the frequency spreads are actually increased. This hints 
that it is also possible to make certain power grid more stable by removing 
some of its links, as long as lines are not removed too destructively with 
respect to a very low threshold. Such power grids may provide a prototype 
for studying the factors that affect the stability of power grids.
Especially, this effect is in analogous to the ``islanding effect'' 
\cite{baldick2008ini,esmaeilian2016}. We will revisit its implication
further in the subsequent subsections.

%%%%%%%%%%%%%%%%%%%%%%%%%%%%%%%%%%%%%%%%%%%%%%%%%%%%%%%%%%%%%%%%%%%%%%%%%
%%%%%%%%%%%%%%%%%%%%%%%%%%%%%%%%%%%%%%%%%%%%%%%%%%%%%%%%%%%%%%%%%%%%%%%%%

%%%%%%%%%%%%%%%%%%%%%%%%%%%%%%%%%%%%%%%%%%%%%%%%%%%%%%%%%%%%%%%%%%%%%%%%%
%%%%%%%%%%%%%%%%%%%%%%%%%%%%%%%%%%%%%%%%%%%%%%%%%%%%%%%%%%%%%%%%%%%%%%%%%

Furthermore, we note that, acting as a damping force, a negative instantaneous
feedback $\alpha=3$ plays the role of slowing down the dynamics of the phase 
order parameter and at the same time suppressing the eventual frequency spread.
A slightly smaller level of decrease in $R(t)$ also happens.
For certain threshold  value, such as the case for $T=0.6$ in 
Fig.~\ref{EUthermcut}, a negative feedback caused nontrivial effect, 
where the frequency spread is dropped by several orders of magnitude. 
Therefore, mechanisms for negative feedback would be desirable in 
designing power grids if achieving a high frequency entrainment is crucial.

Now that we have learned the overall picture for the evolution processes
during thermalization and after carrying out one line cut, in the next 
two subsections, we quantify the changes induced by one line cut
for the US-HV network and EU-HV network, separately.

%%%%%%%%%%%%%%%%%%%%%%%%%%%%%%%%%%%%%%%%%%%%%%%%%%%%%%%%%%%%%%%%%%%%%%%%%
\subsection{US-HV power-grid results \label{sec:3C}}
%%%%%%%%%%%%%%%%%%%%%%%%%%%%%%%%%%%%%%%%%%%%%%%%%%%%%%%%%%%%%%%%%%%%%%%%

We have repeated the analysis performed in~\cite{POWcikk} for 
the US-HV network with the difference that now we do not allow line
cuts in the thermalization, as discussed in the previous subsection. 
This is a more realistic approach, as we avoid failures in the grid 
by a sudden restart. In reality such restart happens by a slow, 
adiabatic procedure, gradually switching on generators and consumers 
by a prescribed power system protocol.

\begin{figure}[ht]
\centering
\includegraphics[width=6.5cm]{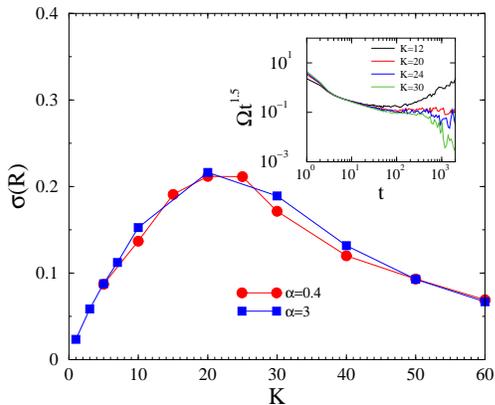}
\caption{\label{betaUSAT} Fluctuations of $R(t \to\infty)$ in case of
the US-HV power grid at the end of the thermalization process.
Both for the normal $\alpha=0.4$ and the $\alpha=3$ dissipation
cases. Partial synchronization transition occurs at $K_c\simeq 22(2)$.
The inset shows the decay of $\Omega(t)$ starting from disordered states
at $\alpha=3$ and for different global couplings, as shown by the legends.
The curves are multiplied by a factor $t^{1.5}$ in order to see the
scaling at $K_c=20$.}
\end{figure}

Generally we started the thermalization processes from fully synchronized 
state: $\theta_i(0)=0$ in order to arrive at higher steady states than 
in case of quenching from random initial conditions. This difference 
comes from the hysteretic behavior of the second-order Kuramoto solution.
As Fig.~\ref{betaUSAT} shows the $R(t \to\infty)$ fluctuations, measured
in the steady state of this thermalization at $t=20000$ exhibit peaks 
as before, published in~\cite{POWcikk}, 
but they occur at a smaller value. We can estimate them as 
$K_c\simeq 22(2)$ for $\alpha=0.4$ and $\alpha=3$ both.

The synchronization transition behavior can also be seen in the $\Omega(t)$ 
decays, starting from disordered states for $\alpha=3$, US-HV, as shown by
the inset of Fig.~\ref{betaUSAT}. 
For $K > K'_c$ the decay is exponentially fast, for $K < K'_c$ 
the curves saturate to finite $\Omega$ values, while at $K'_c=20$ 
a PL decay occurs, characterized by the exponent
$\tau \simeq 1.5$, as one can read off from the inset of 
Fig.~\ref{betaUSAT}. It is interesting to see that $K'_c\approx K_c$ on
the US-HV network, but we should see later that is not always the case.

In~\cite{POWcikk} $K_c\simeq 60(10)$ was published in case of $\alpha=3$,
which is reasonable, since in the thermalization process line cuts were 
allowed, thus higher coupling was needed to stabilize the synchronization.
%%%%%%%%%%%%%%%%%%%%%%%%%%%%%%%%%%%%%%%%%%%%%%%%%%%%%%%%%%%%%%%%%%%%%%%%%
%%%%%%%%%%%%%%%%%%%%%%%%%%%%%%%%%%%%%%%%%%%%%%%%%%%%%%%%%%%%%%%%%%%%%%%%%
Note that in the case of the frequency spread order parameter
we don't find a peak in $\sigma(\Omega(t \to\infty))$, unlike
for the first-order Kuramoto solutions~\cite{Flycikk}, 
thus $\Omega$ does not become more chaotic at the transition.

Yet another advantage of the above-prescribed thermalization process is 
that one can initiate failure cascades by a single line 
or node removal, unlike by the procedures followed in~\cite{Powfailcikk}. 
This alters the steady state values and the synchronization transition 
peaks as compared the results published in~\cite{Powfailcikk}. 

\begin{figure}[ht]
\centering
\includegraphics[width=6.5cm]{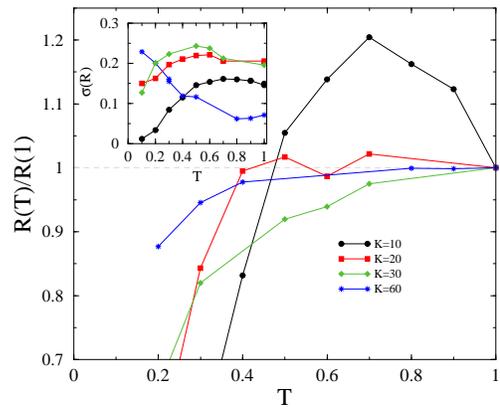}
\caption{\label{UScomp04} Relative change of the steady state
Kuramoto order parameter at $\alpha=0.4$ as the consequence of
the cascade in the US-HV power grid.
The $R(T=1)$ values are set to be the reference points and 
$R(T)/R(T=1)$ is plotted. The gray dashed line marks the baseline
for the emergence of islanding effects. 
The inset shows $\sigma(R(T,K))$. }
\end{figure}

In the previous subsection, we observed that for some intermediate $T$ 
values the order parameter $R$ may even increase following 
one line cut, despite the fact that cascade line failures may follow. To
scrutinize this intriguing observation more closely, the relative change 
of the steady state $R(T)/R(T=1)$ values, following the cascade,
are shown in Fig.~\ref{UScomp04}. Islanding effects occur if $K$ is
lower than the the thermalization value $K_c\simeq 20$ for
$T\gtrsim 0.5$, which is maximal at $T_c\simeq 0.7$ 
(see the inset of Fig.~\ref{UScomp04}).
One may understand this by noting that, on the one hand, cascade line 
failures could bring about disturbances and render the system desynchronized, 
and on the other hand, however, with quite a proportion of the lines 
being removed, the network gradually becomes less and less connected and 
eventually prevent failures from being spread further,
allowing an even higher synchronization level similar to the
``islanding effect''~\cite{baldick2008ini,esmaeilian2016}. The 
observed islanding effect is thus a consequence of the delicate 
competition of these two factors, so that this effect is peaked at
$T_c$, where the increased fluctuations most efficiently do 
the job (cf.~Fig.~\ref{EUcomp04} for a stronger justification).
However, by tuning a system into a very vulnerable state with 
a very small $T$ value, the system becomes quite unstable
and failures will always prevail.
The system is then tremendously desynchronized. 
In the inset of Fig.~\ref{UScomp04} we also show the $\sigma(R(T,K))$ values
for different $K$ and $T$ control parameters. One can observe a smooth
transition for the peaks, suggesting $T_c\to 0$ as we increase $K$.
  
%%%%%%%%%%%%%%%%%%%%%%%%%%%%%%%%%%%%%%%%%%%%%%%%%%%%%%%%%%%%%%%%%%%%%%%%%
%%%%%%%%%%%%%%%%%%%%%%%%%%%%%%%%%%%%%%%%%%%%%%%%%%%%%%%%%%%%%%%%%%%%%%%%%

\begin{figure}[ht]
\centering
\includegraphics[width=6.5cm]{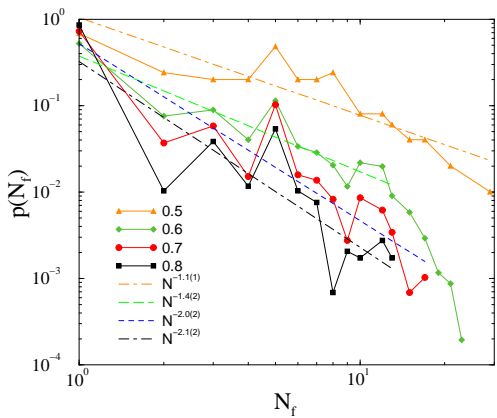}
\caption{\label{USPL30} 
Probability distribution of line failures for $K = 30$, $\alpha=0.4$ and
different thresholds marked by the legends in case of the US-HV power grid.
Dashed lines show power-law fit for the scaling region, determined by visual
inspection.}
\end{figure}

\begin{figure}[ht]
\centering
\includegraphics[width=6.5cm]{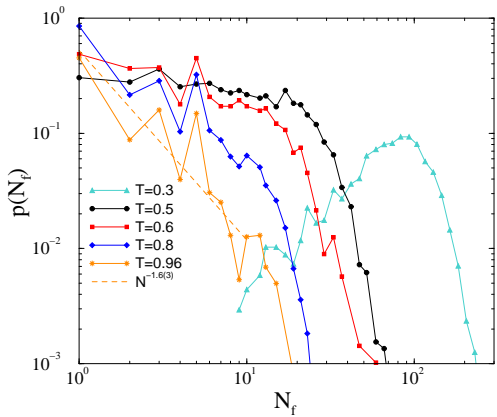}
\caption{\label{USPL20} 
Probability distribution of line failures for $K = 20$, $\alpha=0.4$ and
different thresholds marked by the legends in case of the US-HV power grid.
DK cascade failures at $T=0.3$ can be observed.}
\end{figure}

The US-HV cascade size distributions $p(N_f)$ published in
\cite{Powfailcikk} remain the same for $K=30$, as demonstrated on 
Fig.~\ref{USPL30}. The estimated exponent values are also in
good agreement with empirical estimations \cite{Car2}.
For $K=20$ we could find $p(N_f)$ with PL tails, at $T \ge 0.9$ only.
Here we could fit for the narrow region $N_f<10$ a PL with
an exponent $\tau=1.6(3)$, as shown in Fig.~\ref{USPL20}.
As we increase $T$ from 0.5, we can see slowly decaying $p(N_f)$
curves, with an exponential cutoff for $N_f > 20$.
For $T < 0.5$ thresholds, bumps in the PDFs with large mean values
appear, suggesting Dragon King (DK) events of cascade failures.
However, these DK bumps do not appear for couplings $K>K_c$. 

Note, the forms of $p(N_f)$ do not change as we increase the 
dissipation/feedback factor from  $\alpha=0.4$ to $\alpha=3$. 
The PDFs are also insensitive for the mode of initial perturbation, 
i.e.~whether an edge or a node removal is committed after the thermalization. 
This kind of invariance was also found in \cite{Powfailcikk}
where even multiple perturbations caused the same kind of cascade size
distribution forms.

%%%%%%%%%%%%%%%%%%%%%%%%%%%%%%%%%%%%%%%%%%%%%%%%%%%%%%%%%%%%%%%%%%%%%%%%%
%%%%%%%%%%%%%%%%%%%%%%%%%%%%%%%%%%%%%%%%%%%%%%%%%%%%%%%%%%%%%%%%%%%%%%%%%
 
%%%%%%%%%%%%%%%%%%%%%%%%%%%%%%%%%%%%%%%%%%%%%%%%%%%%%%%%%%%%%%%%%%%%%%%%%
%%%%%%%%%%%%%%%%%%%%%%%%%%%%%%%%%%%%%%%%%%%%%%%%%%%%%%%%%%%%%%%%%%%%%%%%%

%%%%%%%%%%%%%%%%%%%%%%%%%%%%%%%%%%%%%%%%%%%%%%%%%%%%%%%%%%%%%%%%%%%%%%%%%
\subsection{EU-HV power-grid results \label{sec:3D}}
%%%%%%%%%%%%%%%%%%%%%%%%%%%%%%%%%%%%%%%%%%%%%%%%%%%%%%%%%%%%%%%%%%%%%%%%

\begin{figure}[ht]
\centering
	\includegraphics[width=0.4\textwidth]{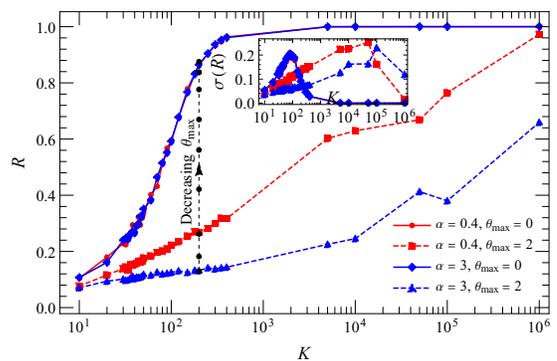}
\caption{\label{EUhyst} The steady state values of the Kuramoto
order parameter for the EU-HV network. Start from 
ordered/disordered initial states causes a hysteresis loop, 
closing at $K \sim 10^6$ for $\alpha=0.4$. 
For $\alpha=3$ we could not reach the closure point as for 
this large value the adaptive solver becomes very slow. 
There exists infinite intermediate branches if we initialize the 
phases as $\theta_i(0) \in (0,\theta_{\mathrm{max}})$ with 
different $\theta_{\mathrm{max}}$ values.
The inset shows $\sigma(R)$, in which the peaks of the solid
curves indicate $K_c\simeq 100$.}
\end{figure}

By determining the steady state values of $R$ after the thermalization 
we can investigate the type of transition as it was done for the US-HV
case. The adaptive solver allowed us to provide precise values even for 
very large $K$ values. Similar to the US-HV grid we again found a smooth 
crossover from desynchronization to partial synchronization (which is 
merely indicated by the $\sigma(R)$ peaks for the US-HV case 
in~Fig.~\ref{betaUSAT}) by increasing the global coupling $K$, 
as shown on Fig.~\ref{EUhyst}.
We can observe a hysteresis, where the upper branch corresponds to a start
from an ordered, while the lower branch to a disordered initial state.
The upper branch is insensitive to $\alpha$, but for very large $K$-s 
the Runge-Kutta-4 algorithm breaks down and provides nonphysical, 
decreasing order parameter values. 
We have also measured the fluctuations of $R$ over many samples. 
The inset of Fig.~\ref{EUhyst} shows that for the upper branch a 
peak occurs at $K\simeq 100$ marking a transition there. 
For random initial states the fluctuation peaks seem to occur at 
much higher couplings, that we could not investigate in detail 
as the adaptive solver slows down substantially for large $K$ values.

%%%%%%%%%%%%%%%%%%%%%%%%%%%%%%%%%%%%%%%%%%%%%%%%%%%%%%%%%%%%%%%%%%%%%%%%%
%%%%%%%%%%%%%%%%%%%%%%%%%%%%%%%%%%%%%%%%%%%%%%%%%%%%%%%%%%%%%%%%%%%%%%%%

\begin{figure}[ht]
\centering
\includegraphics[width=6.5cm]{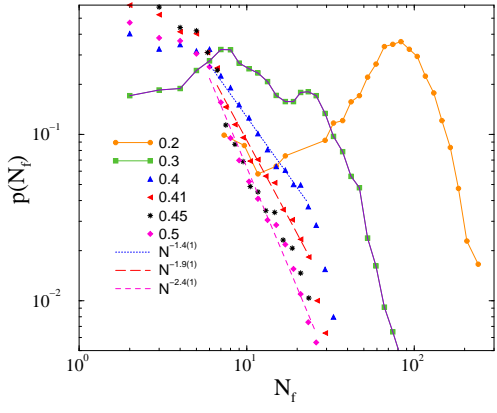}
\caption{Probability distribution of line failures for different 
thresholds for $K=60$ shown in the legends in case of the EU-HV power grid.
Dashed lines show power-law fits for the scaling region, determined by visual 
inspection. One can observe DK bumps for low thresholds.}
\label{EUPL60}
\end{figure}

We have also investigated the cascade size distributions for $\alpha=0.4, 3$, 
$K = 20,60$, and $80$, i.e.~near the synchronization transition
point $K_c\simeq 100$. 
For the low $K=20$ we could only find log-normal like PDFs, with increasing
mean values as $T$ is decreased; see Fig.~\ref{EUPL20} in the Appendix, in
hich we also demonstrate the invarience of $p(N_f)$ with respect to an
initial line or node removal.
Still below the synchronization transition point $K_c$, at $K=60$ 
one finds very few cascade line failures for $T > 0.5$, PDFs 
with continuously changing PL tails for $0.5 \le T \le 0.4$ and 
log-normal like distributions for $T < 0.5$ (see Fig.~\ref{EUPL60}). 
The small threshold distributions resemble to the so called 
``Dragon King'' events described in Refs.~\cite{Sorn-Uui,Gur-Sou}, 
while in the $0.5 \le T \le 0.4$ region, the continuously changing 
exponents $1.4 \le \tau \le 2.4$ suggest Griffihts effects as the 
consequence of the heterogeneity. Note, that Griffihts effects
may also occur near mixed order transitions, where the steady state
values jump, but dynamical scaling persists~\cite{OdorSim21}.

\begin{figure}[ht]
\centering
\includegraphics[width=6.5cm]{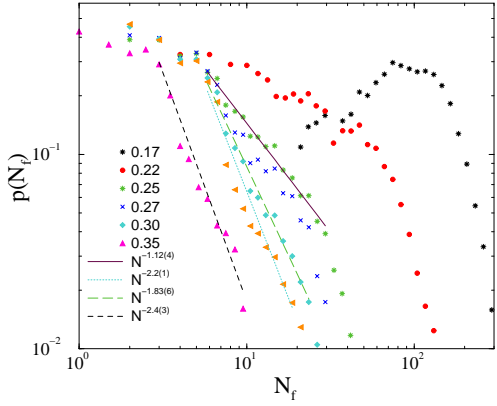}
\caption{Probability distribution of line failures for different thresholds
for $K=80$ shown in the legends in case of the EU-HV power grid.
Dashed lines show power-law fits for the scaling region determined by visual 
inspection. One can observe DK bumps for low thresholds.
\label{EUPL80}}
\end{figure}

Near the transition point, at $K=80$ the situation is different 
as shown on Fig.~\ref{EUPL80}. 
Here PL tails emerge in the region $0.33 \le T \le 0.46$, 
which could be fitted by the exponents $1.3 \le \tau < 3.0$, but 
part of these exponent values are larger than the empirical 
data and simulations of Refs.~\cite{Car,Car2}, $1.3 \le \tau \le 2$
obtained for many countries.
As these PL decays are very steep and occur for $N_f < 20$ only, 
it is hard to differentiate them from the tail of a Gaussian
distribution. For $T < 0.22$ thresholds we can find the emergence
of bumps, suggesting DK events.

As thermal noise may occur in real systems, which has not been
considered up to now, but fluctuations, which came from the nonlinear 
chaoticity, we have also investigated the effect of an additional 
Gaussian noise. Without showing the preliminary results, we
remark that the zero centered annealed noise, with unit variance
did not modify the shape of the PDF, but shifted the noiseless
$T=0.33$ result to $T=0.56$. 
Further analysis of the thermal noise is in progress, using
the Euler-Maruyama adaptive solver, that can avoid the problem with
Bulirsch-Stoer and the inaccuracies of Runge-Kutta-4 at large couplings.

%%%%%%%%%%%%%%%%%%%%%%%%%%%%%%%%%%%%%%%%%%%%%%%%%%%%%%%%%%%%%%%%%%%%%%%%%

%%%%%%%%%%%%%%%%%%%%%%%%%%%%%%%%%%%%%%%%%%%%%%%%%%%%%%%%%%%%%%%%%%%%%%%%%

%%%%%%%%%%%%%%%%%%%%%%%%%%%%%%%%%%%%%%%%%%%%%%%%%%%%%%%%%%%%%%%%%%%%%%%%%

%%%%%%%%%%%%%%%%%%%%%%%%%%%%%%%%%%%%%%%%%%%%%%%%%%%%%%%%%%%%%%%%%%%%%%%%%

\begin{figure}[ht]
\centering
\includegraphics[width=6.5cm]{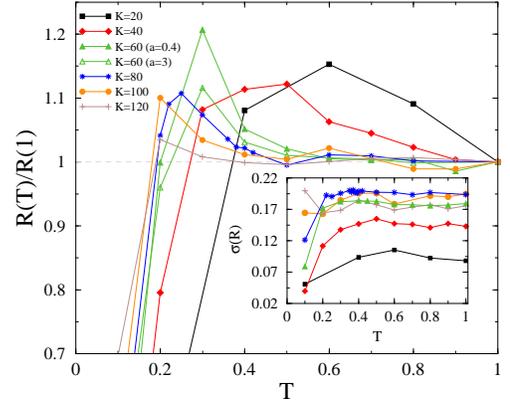}
\caption{\label{EUcomp04} Relative change of the steady state
Kuramoto order parameter at $\alpha=0.4$ the consequence of 
the cascade in the EU-HV grid. The $R(T=1)$ values are set to be the 
reference points and $R(T)/R(T=1)$ is plotted. 
The gray dashed line marks the baseline for the emergence of islanding 
effects. The inset shows $\sigma(R(T,K))$.}
\end{figure}

We have compared the stability of steady states of the US-HV and
EU-HV systems as the consequence of the failure cascades
by measuring $R(t \to\infty)$.
Since the sizes and dimensions of the networks differ, we determined the
relative change of $R$ (Fig.~\ref{EUcomp04}) as in the case of US-HV
network. Again, we can see the ``islanding'' stabilization effect
on the networks after the cascade failures, known in the power-grid 
engineering literature \cite{baldick2008ini,esmaeilian2016}. 
As mentioned in the last subsection, this happens near $T_c(K)$, 
as one can read off from the inset of Fig~\ref{EUcomp04},
but gradually disappears for large $K$ above $K_c$.

Note, that the maxima of these curves also agree with the $T$ values 
by which the $p(N_f)$ decays exhibit power-law decay behavior. 
Hence, the system displays some nontrivial critical dynamics in the 
vicinity of criticality $(K_c, T_c)$. We can also see that the increase 
of the dissipation factor from $\alpha=0.4$ to $\alpha=0.3$ does 
not change a lot, but enhances the islanding effect; compare the 
$K=60$ curves in Fig.~\ref{EUcomp04}.

%%%%%%%%%%%%%%%%%%%%%%%%%%%%%%%%%%%%%%%%%%%%%%%%%%%%%%%%%%%%%%%%%%%%%%%%%
%%%%%%%%%%%%%%%%%%%%%%%%%%%%%%%%%%%%%%%%%%%%%%%%%%%%%%%%%%%%%%%%%%%%%%%%%

%%%%%%%%%%%%%%%%%%%%%%%%%%%%%%%%%%%%%%%%%%%%%%%%%%%%%%%%%%%%%%%%%%%%%%%%%
\section{Local Kuramoto results \label{sec:4}}
%%%%%%%%%%%%%%%%%%%%%%%%%%%%%%%%%%%%%%%%%%%%%%%%%%%%%%%%%%%%%%%%%%%%%%%%%

\begin{figure*}[ht]
\centering
\includegraphics[width=0.96\textwidth]{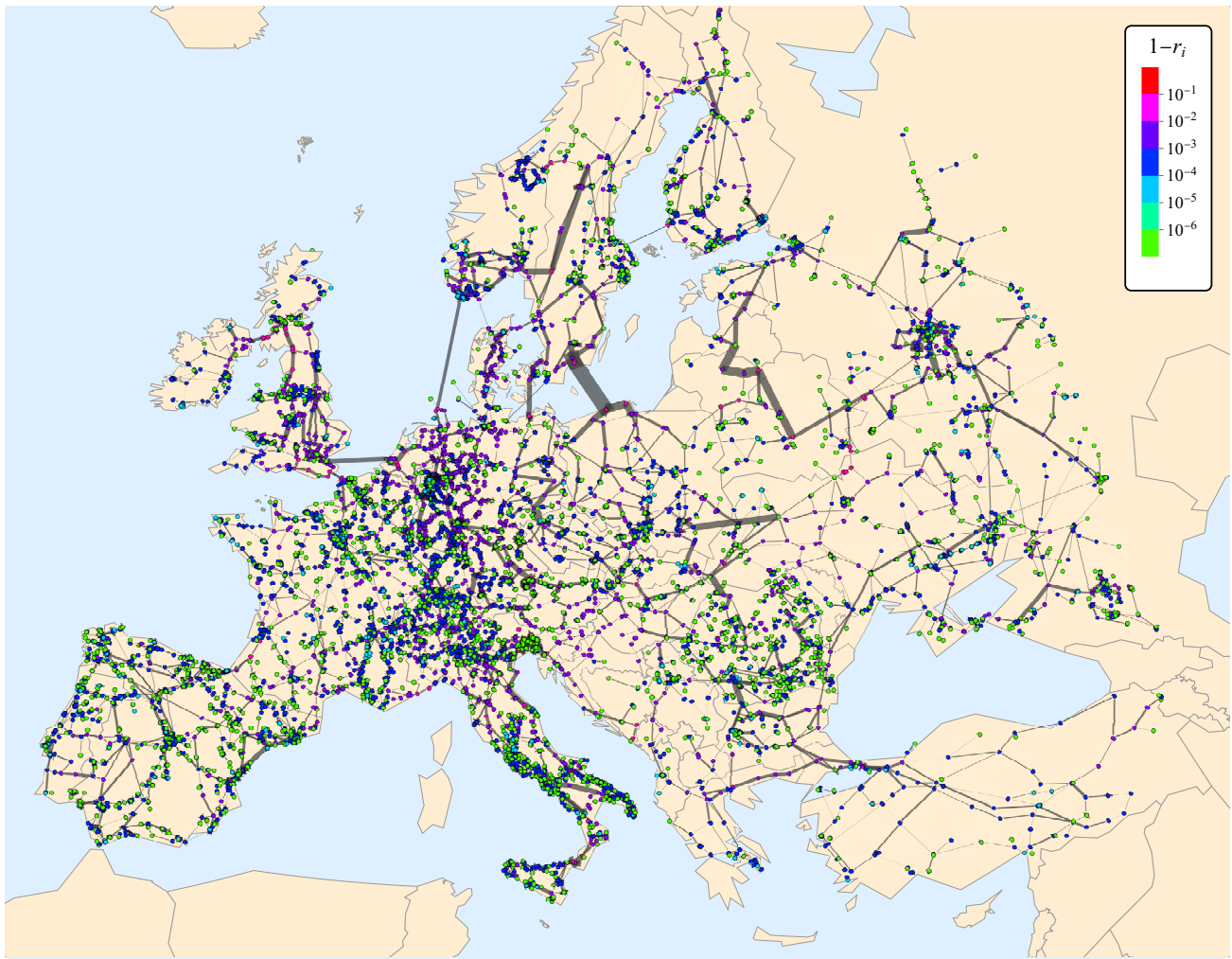}
\caption{\label{EU80T1} Local Kuramoto results encoded by the color
map as $1-r_i$. Red corresponds low local synchronization,
green to high synchronization. The width of gray edges is
proportional to the amplitude of the power flow.}
\end{figure*}

To investigate the heterogeneity further we also measured the local
Kuramoto order parameter, defined as the partial sum of phases 
for the neighbors of node $i$
\begin{equation}\label{LCO}
	r_i(t)= \frac{1}{N_{\mathrm{i.neigh}}}
		\left|\sum_j^{N_{\mathrm{i.neigh}}}  A_{ij} e^{i
	\theta_j(t)}\right|  \ ,
\end{equation}
and the power-flows $F_{ij}$ defined in Eq.~\eqref{Fij}.
This local Kuramoto measure was firstly suggested by Restrepo
\textit{et al.} \cite{restrepo2005,schroder2017} to quantify the local 
synchronization of nodes, which allows us to follow the synchronization 
process by mapping the solutions on the geographical map.
One example for the EU-HV grid is shown in Fig.~\ref{EU80T1}, 
which is the result of thermalization after averaging over 1000 samples, 
initiated by random self-frequencies at $K=80$ and $\alpha=0.4$,
rendering the grid close to the transition point. 
The map reveals strongly synchronized (green)
regions as well as some weakly synchronized ones, especially near sea
cable connections, where the power-flow is also maximal. Hence, it is
quite evident that there are many heterogeneous, rare regions in the
EU-HV grid.

%%%%%%%%%%%%%%%%%%%%%%%%%%%%%%%%%%%%%%%%%%%%%%%%%%%%%%%%%%%%%%%%%%%%%%%%
\section{Conclusions\label{sec:5}}
%%%%%%%%%%%%%%%%%%%%%%%%%%%%%%%%%%%%%%%%%%%%%%%%%%%%%%%%%%%%%%%%%%%%%%%%

Energy security has becoming an extremely important issue these days.
The vulnerability of power grids depend very much on the network
topology, which has been investigated in our numerical study by
comparing the western US and the EU high-voltage grids.
The traditional electric swing equations, describing the power-flow 
among interconnected rotating generators and consumers (machines)
is equivalent to a set of second-order Kuramoto equations, which
have been solved numerically on these networks. 
Our analysis has two parts, a thermalization regime, by which we 
create stable steady state conditions and a ``line-cut'' regime, 
where we allowed dynamical cascade failure events, by removing edges
from the graph, whenever the power flows exceed certain 
threshold values.

The time dependent behavior in the thermalization regime had been
followed by analyzing the Kuramoto order parameter of phases as well
as by the frequency spread. We used a simplified scenario, in which
self-frequencies of the nodes were chosen randomly from a zero centered
Gaussian distribution, thus we neglected the power in/out magnitudes
of the vertices and their inertia as well as the impedance 
of the lines. Instead we focused on the effects of dissipation, which
can also describe instantaneous feedback.

We compared the frequency spread behavior in the large coupling limit 
of the solutions with those of the linearized equations and showed 
that they agree in the case of coherent initial state using $d=2,3$ 
lattices. This behavior is asymptotically the same as that of the 
massless Kuramoto equation: $\Omega(t) \sim t^{-d/2}$. 
However, starting from randomly distributed oscillators we observed
faster decays with PL regions after a short initial slip and 
before finite-size cutoffs. We also found this behavior at the 
synchronization transitions for the US and EU grids, 
suggesting that when the linear approximation breaks down due
to the fluctuations, coming by large chaoticity at the
transition or by initial conditions, non-trivial PL decay can
be observed for intermediate times in finite systems. 

It is important to note that for the frequency order parameter we 
expect real phase transitions for $d \ge 2$, which was the case in 
our networks. For $R$ we expect crossovers in case of $d < 4$, which
could be identified by the fluctuation peaks, agreeing with the 
phase order parameter transition data. However, this crossover point
may not coincide with the frequency phase transition point value
for the following reason. The crossover point $K_c\to\infty$ with 
the system size, while a real phase transition point should remain a
finite value in the thermodynamic limit. In case of US-HV we found
a coincidence $K'_c \approx K_c$ regardless of the initial conditions 
(cf.~Fig.~\ref{betaUSAT} and Fig.~\ref{omegatr} in the Appendix), 
while for the EU-HV the the phase transition of $\Omega$
happens at $K'_c\simeq 44$ as compared to $K_c\simeq 100$ crossover
point of the phases (cf.~Fig.~\ref{EUhyst} and Fig.~\ref{omegatr}
in the Appendix).

The dissipation/feedback factor had mainly delaying effects on the
$R(t)$, but increasing $\alpha$ slightly increased $R(t \to \infty)$ 
and drastically decreased $\Omega(t\to \infty)$. In general the $\Omega$
order parameter proved to be a more sensitive measure of the 
synchronization than $R$ .

Following the thermalization we induced failure cascades, by cutting
a line from the network. We measured the probability distributions 
of the line failures, which occurred by the overloads of lines 
during the power redistribution process, dictated by the 
second-order Kuramoto dynamics. In the US-HV case, the PDFs 
of the cascade line cuts were log-normal, for small avalanches
in case of high $T$ and $K$ values. For the control parameter
regions, where the fluctuations of $R$ showed a peak, the PDFs 
exhibited PL tails, with continuously changing exponents around 
$\tau \simeq 1.9$,  which agrees with historically observed blackout 
exponent vales. 
For the EU case we also found the occurrence of steeper PL-s, 
characterized by $\tau \simeq 3$, in contrast to the US-HV case,
but as the observed scaling region is narrow, possibly due to finite sizes, 
it is hard to distinguish PL-s from log-normal.
In any case this is in agreement with the topological analysis 
result by ~\cite{CasTh} according to which the marginally fragile EU 
network might show scale-free cascade distributions as well.

For small $T$ and $K$ parameters non-zero centered PDF ``bumps'' 
could be observed, which can be explained by the first-order 
transition behavior of the second-order Kuramoto model and can 
be called Dragon Kings following the literature.

In the PL scaling regions we found that the remaining networks after 
the cascade could exhibit an increase of synchronization. 
This kind of ``islanding'' stabilization is also known in the literature 
as a mean to stabilize power-grids via truncation.
The detailed phase transition analysis leads to our most 
important conclusion for blackout engineering: 
although ``islanding'' stabilization does not work for $K > K_c$, 
it is just the best near the transition point $T_c(K)$, 
thus self-organization of power systems provides an advantage, 
as in many other cases, like in neural systems.

Our results suggest that the second-order Kuramoto model
exhibits a mixed-order transition in the sense that
the steady state order parameter $R(t\to\infty)$ displays a hysteresis
and the frequency order parameter $\Omega(t\to\infty)$ jumps rapidly 
at the transition point, with $\Omega(t)$ exhibiting PL tails there.
Furthermore, in case of allowing line failures the PDFs of the 
avalanches also show PL tails. As the consequence of the strong 
heterogeneity Griffihts phase like frustrated synchronization 
can also be observed. This is allowed, because the graph 
dimensions are below $d_c=4$ and the networks possess 
modular structure, which enhances rare region effects.

It is important to note that these PL regions are robust for
changes in the networks and the initial perturbations details.
The exponents are also invariant for the value of $\alpha$, 
but the width of the distribution shrinks, so instantaneous
feedback can stabilize or delay cascades, but does not affect 
the tail behavior of the PDFs.
 
We have been investigating the heterogeneity in more detail by measuring
the local order parameters. As a preliminary result we show how the 
stronger connected graphical regions are reflected by higher local 
synchronization.
 
\section*{Appendix}

In this Appendix we first show cascade failure distribution 
results for the EU grid in case of a very low global coupling: 
$K=20$ as compared to the transition point: $K_c\simeq 100$.
Although the sizes of the avalanches are large, we can't see 
scale-free behavior up to $T = 0.8$. 
The graph also shows invariance of PDFs
with respect to an initial node or line failure.

%%%%%%%%%%%%%%%%%%%%%%%%%%%%%%%%%%%%%%%%%%%%%%%%%%%%%%%%%%%%%%%%%%%%%%%%%
\begin{figure}[ht]
\centering
\includegraphics[width=6.5cm]
{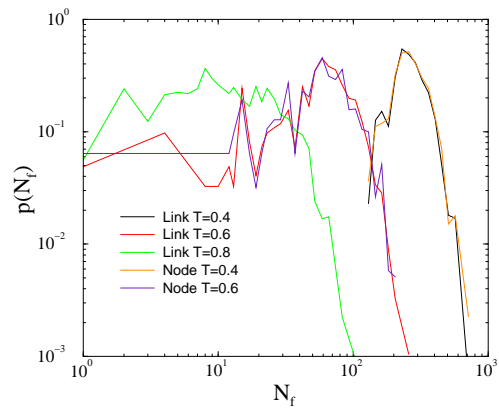}
\caption{Probability distribution of line failures for different thresholds
for $K=20$ shown in the legends in case of the EU-HV power grid.
\label{EUPL20}}
\end{figure}
%%%%%%%%%%%%%%%%%%%%%%%%%%%%%%%%%%%%%%%%%%%%%%%%%%%%%%%%%%%%%%%%%%%%%%%%%

%%%%%%%%%%%%%%%%%%%%%%%%%%%%%%%%%%%%%%%%%%%%%%%%%%%%%%%%%%%%%%%%%%%%%%%%%
\begin{figure*}[ht]
\centering
\begin{tabular}{cc}
\includegraphics[height=6.cm]{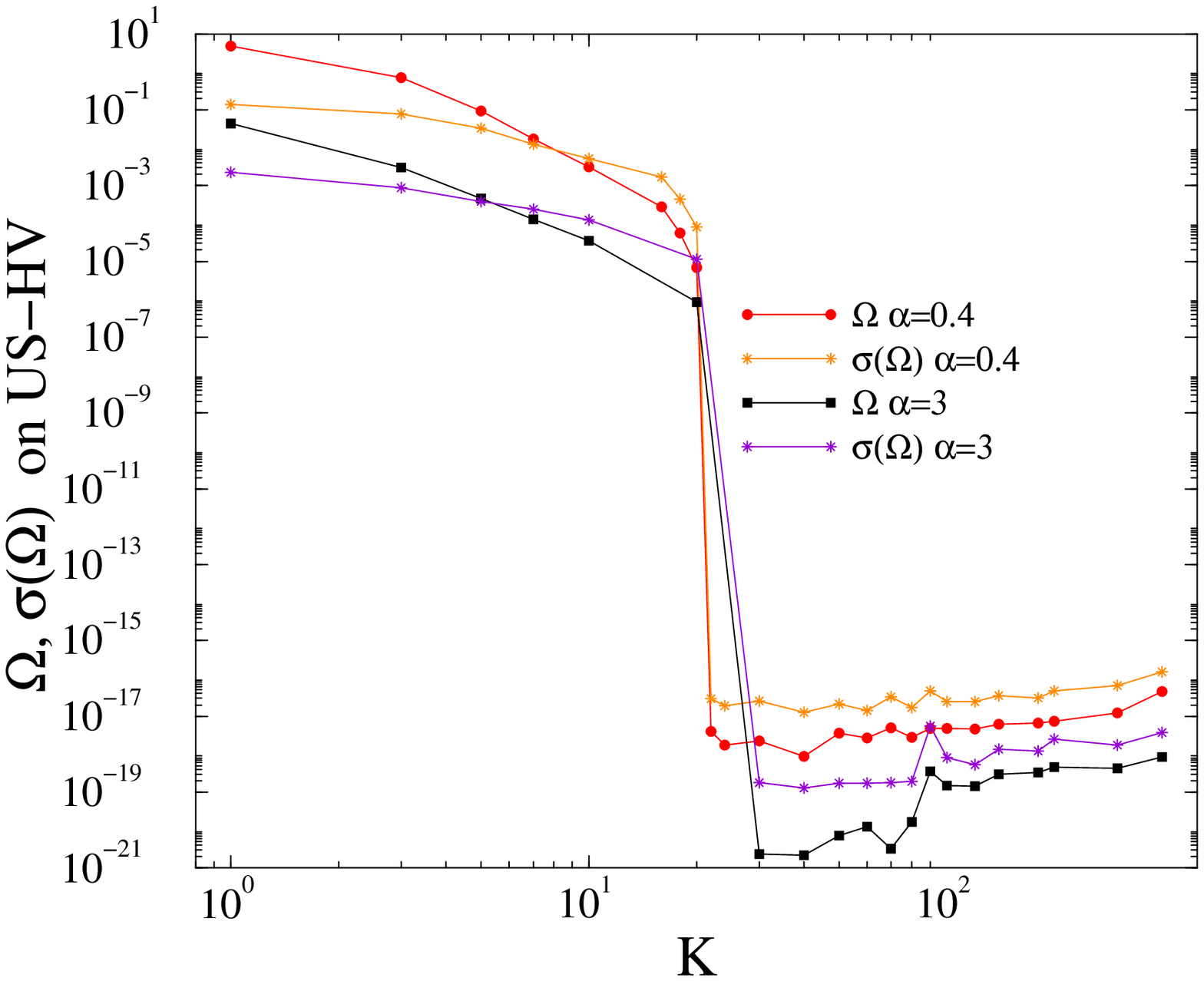} & \ \ \ \ 
\includegraphics[height=6.cm]{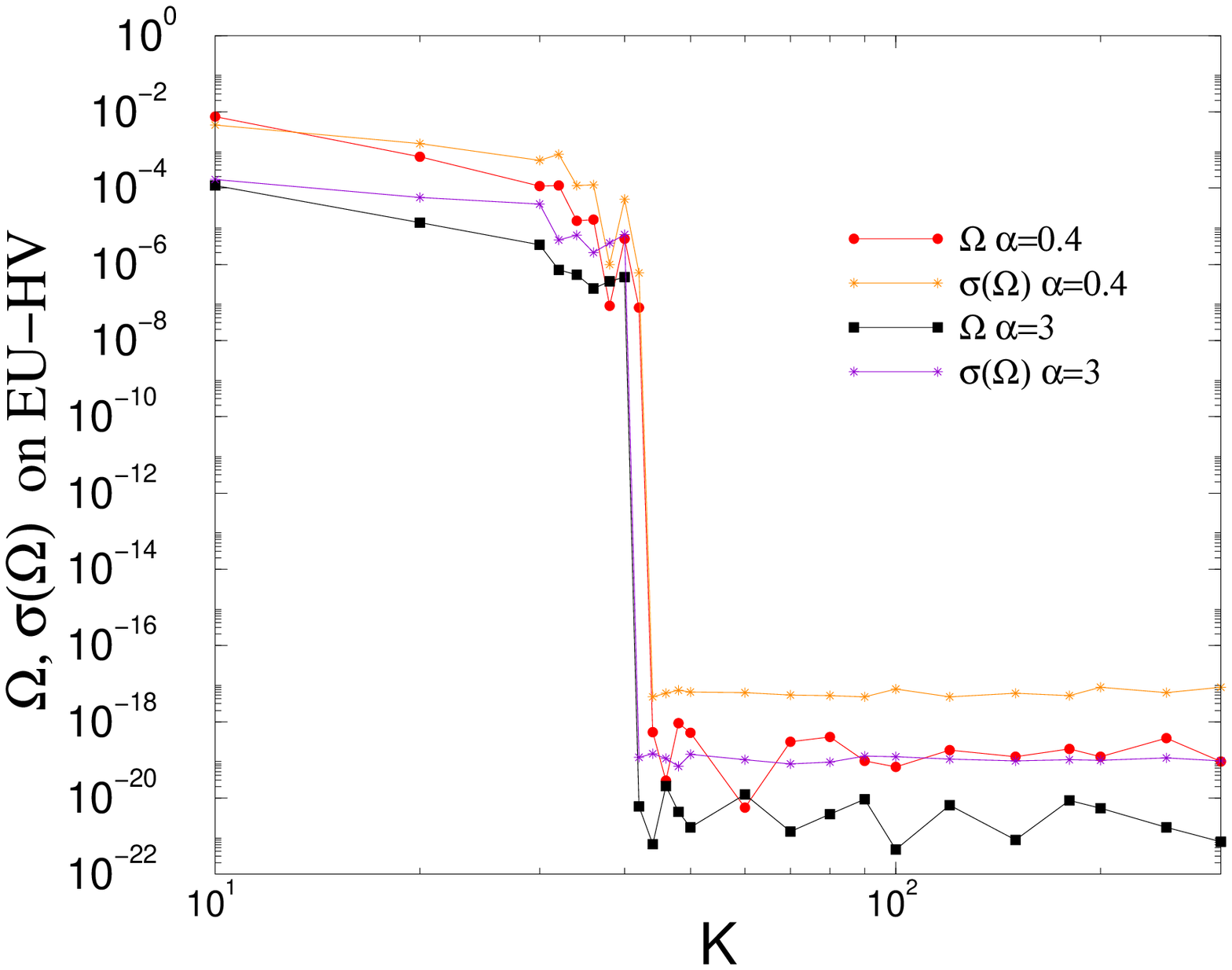} \\
\end{tabular}
\caption{The stationary frequency spread and the corresponding standard 
	deviation show that the frequency entrainment has a 
	transition point at $K'_c\simeq 20$ for the US-HV grid (Left) and
	at $K'_c\simeq 20$ for the EU-HV grid (Right). This transition
	point coincides with $K_c$ for the US-HV case but differs from that
	of the EU-HV case, as estimated from the peaks of $\sigma(R)$; 
	cf.~Figs.~\ref{betaUSAT} and \ref{EUhyst}.}
\label{omegatr}
\end{figure*}
%%%%%%%%%%%%%%%%%%%%%%%%%%%%%%%%%%%%%%%%%%%%%%%%%%%%%%%%%%%%%%%%%%%%%%%%%

In addition, Fig.~\ref{omegatr} shows that there
is also a transition in the frequency spread. Since the lower critical
dimension for frequency entrainment and phase order parameter are
$2$ and $4$, respectively, in $2\le d < 4$, the frequency
entrainment transition point $K'_c$ is not necessarily identical with
the phase order crossover point $K_c$ \cite{HPCE}. Even though the 
transition point $K'_c$ for the US-HV case almost coincides with $K_c$ 
we see that $K'_c\simeq 44$ is rather different from $K_c\simeq 100$ in
the EU-HV case.

\acknowledgments{

We thank R\'obert Juh\'asz and Lilla Barancsuk for the useful comments.
Support from the advantaged ELKH grant and the Hungarian National Research, 
Development and Innovation Office NKFIH (K109577) is acknowledged.
%The VEKOP-2.3.2-16-2016-00011 grant is supported by the European Structural
%and Investment Funds jointly financed by the European Commission and the
%Hungarian Government.
Most of the numerical work was done on KIFU supercomputers of Hungary.
}

%%%%%%%%%%%%%%%%%%%%%%%%%%%%%%%%%%%%%%%%%%%%%%%%%%%%%%%%%%%%%%%%%%%%%%%%
%\bibliography{bib}
%%%%%%%%%%%%%%%%%%%%%%%%%%%%%%%%%%%%%%%%%%%%%%%%%%%%%%%%%%%%%%%%%%%%%%%%

\end{document}